# Solitons in an extended nonlinear Schrödinger equation with third-order dispersion and pseudo-Raman effect

A.V. Aseeva, L.G. Blyakhman, E.M. Gromov, V.V. Tyutin

National Research University Higher School of Economics, Nizhny Novgorod 603155, Russia

**Abstract**

Dynamics of solitons is considered in an extended nonlinear Schrödinger equation, including a pseudo-stimulated-Raman-scattering (pseudo-SRS) term (scattering on damping low-frequency waves, third-order dispersion (TOD) and inhomogeneity of the spatial second-order dispersion (SOD). It is shown that wave-number downshift by the pseudo-SRS may be compensated by upshift provided by spatially increasing SOD with taking into account TOD. The equilibrium state is stable for positive parameter of TOD and unstable for negative one. The analytical solutions are verified by comparison with numerical results

## 1. Introduction

The great interest to the dynamics of solitons is motivated by their ability to travel long distances keeping the shape and transferring the energy and information. Soliton solutions are relevant to nonlinear models in various areas of physics which deal with the propagation of intensive wave fields in dispersive media: optical pulses and beams in fibers and spatial waveguides, electromagnetic waves in plasma, surface waves on deep water, etc. [1-7].

Dynamics of long high-frequency (HF) wave packets is described by the second-order nonlinear dispersive wave theory. The fundamental equation of the theory is the nonlinear Schrödinger equation (NLSE) [8,9], which includes the second-order dispersion (SOD) and cubic nonlinearity. Soliton solutions in this case arise as a result of the balance between the dispersive stretch and nonlinear compression of wave packets.

To solve many applied problems there is the necessary decreasing of solitons`s space size. Such decreasing is accompanied, as usually, by stimulated scattering on low-frequency (LF) media perturbations. To this time stimulated scattering on spatially homogeneous LF time modes (stimulated Raman scattering (SRS)) was considered in details [1]. SRS is described in extended NLSE by term with time delay of nonlinear kerr response. For localized nonlinear wave packets (solitons), the SRS gives rise to the downshift of the soliton frequency [10]. The compensation of the SRS was studied to this time distally [10-21].

For a series media the propagation of short solitons is accompanied by arising of damping LF waves. These LF modes are internal waves in the stratified fluid and ion-sound waves in the plasma. Model for describing of stimulated scattering of HF waves on damping LF waves, named as pseudo-stimulated-Raman-scattering (pseudo-SRS), was proposed in [22-24]. Taking into account the wave factor of stimulated LF perturbations significantly varieties the dynamics of short HF solitons. The pseudo-SRS leads to the self-wavenumber downshift, similar to what is well known in the temporal domain [1,10-21]. The model equation elaborated in [22-24] also included smooth spatial variation of the SOD, accounted for by a spatially decreasing SOD coefficient, which leads to an increase of the soliton's wave-number, making it possible to compensate the effect of the pseudo-SRS on the soliton by the spatially inhomogeneous SOD. The equilibrium between the pseudo-SRS and decreasing SOD gives rise to stabilization of the soliton's wave-number spectrum. However, the consideration was carried out in disregard of the TOD.

In this work the soliton dynamics is considered in the frame of extended NLSE with a pseudo-SRS, decreasing dispersion and with taking into account nonlinear dispersion. Shows that equilibrium state between the pseudo-SRS and decreasing SOD is stable focus for positive TOD, and unstable focus for negative TOD.

## 2. The basic equation and integral relations

Let's consider the dynamics of the HF wave field $U(\xi,t)\exp(i\omega t - i\kappa\xi)$ in the frame of extension NSE with pseudo-SRS, TOD and inhomogeneous SOD:

$$2i\frac{\partial U}{\partial t} + \frac{\partial}{\partial \xi}\left[q(\xi)\frac{\partial U}{\partial \xi}\right] + 2U|U|^2 + i\gamma\frac{\partial^3 U}{\partial \xi^3} + \mu U \frac{\partial(|U|^2)}{\partial \xi} = 0, \tag{1}$$

where $q(\xi)$ is the SOD, $\mu$ is the pseudo-SRS, $\gamma$ is the TOD. Equation (1) with zero conditions on infinity $U|_{\xi\to\pm\infty} \to 0$ has the following integrals:

$$\frac{dN}{dt} \equiv \frac{d}{dt}\int_{-\infty}^{+\infty}|U|^2 d\xi = 0, \tag{2}$$

$$2\frac{d}{dt}\int_{-\infty}^{+\infty} K|U|^2 d\xi = -\mu\int_{-\infty}^{\infty}\left[\frac{\partial(|U|^2)}{\partial \xi}\right]^2 d\xi - \int_{-\infty}^{\infty}\frac{dq}{d\xi}\left|\frac{\partial U}{\partial \xi}\right|^2 d\xi, \tag{3}$$

$$N\frac{d\overline{\xi}}{dt} \equiv \frac{d}{dt}\int_{-\infty}^{\infty}\xi|U|^2 d\xi = \int_{-\infty}^{+\infty} qK|U|^2 d\xi - \frac{3}{2}\gamma\int_{-\infty}^{\infty}\left|\frac{\partial U}{\partial \xi}\right|^2 d\xi, \tag{4}$$

where $U \equiv |U|\exp(i\phi)$, $K \equiv \partial\phi/\partial\xi$ is the local wave-number of wave packet.

## 3. Analytical results

For analysis of the system (2)-(4) let use the adiabatic approximation, presenting solution in sech-like form

$$U(\xi,t) = A(t)\mathrm{sech}\left[\frac{\xi - \overline{\xi}(t)}{\Delta(t)}\right]\exp\left[i\int K(\xi,t)d\xi - \frac{i}{2}\int A^2(t)dt\right],$$

$$K(\xi,t) = k(t) + \frac{3}{2}\frac{\gamma}{q(\overline{\xi})}\tanh^2\left[\frac{\xi - \overline{\xi}(t)}{\Delta(t)}\right] + \frac{\gamma}{q(\overline{\xi})}\tanh\left[\frac{\xi - \overline{\xi}(t)}{\Delta(t)}\right], \tag{5}$$

where $\Delta(t) = \sqrt{q(\bar{\xi}(t))}/A(t)$, $\Omega(t) = A^2(t)/2$, $A^2(t)\Delta(t) = \text{const}$. With taking into account (2), (5) and system (2)-(4) is reduced to:

$$2\frac{dk}{dt} = -\frac{8q_0^2 A_0^4 \mu}{15q^3} - \frac{q' q_0 A_0^2}{3q^2} - q'k^2 + \frac{4q' q_0 \gamma A_0^2 k}{q^3}, \quad \frac{d\bar{\xi}}{dt} = qk, \tag{6}$$

where $q_0 = q(0)$, $A_0 = A(0)$, $q'(\bar{\xi}) = (dq/d\xi)_{\bar{\xi}}$. System (6) give rise to an obvious equilibrium state (alias fixed point, FP): $8q_0 A_0^2 \mu = -5q'(\bar{\xi}_*)q(\bar{\xi}_*)$, $k_* = 0$. In particular, for

$$\mu = \mu_* \equiv -5q'(0)/(8A_0^2) \tag{7}$$

the FP corresponds to initial soliton parameters: $\bar{\xi} = 0$, $k = 0$. For $\mu \neq \mu_*$ soliton`s parameters are time-varying. To analyze the evolution around the FP, we assume linearly decreasing SOD, $q' = \text{const} < 0$, and rescale the variables by defining $\tau \equiv -tq'A_0/\sqrt{3q_0}$, $y \equiv k\sqrt{3q_0}/A_0$ and $n = q(\bar{\xi})/q_0$. Then system (6) is reduced to

$$2\frac{dy}{d\tau} = -\frac{\lambda}{n^3} + \frac{1}{n^2} + y^2 - I\frac{y}{n^3}, \quad \frac{dn}{d\tau} = -ny, \tag{8}$$

where $\lambda \equiv -8\mu A_0^2/(5q') \equiv \mu/\mu_*$, $I \equiv 4\sqrt{3}\gamma A_0/\sqrt{q_0^3}$. The FP of the Eqs. (8) in rescale variables is $y_* = 0, n_* = \lambda$. For $I > 0$ the FP is the stable focus, $I = 0$: center, $I < 0$: unstable focus. Trajectories on plane $(y, n)$ found from Eqs. (8) with the initial conditions $y_0 = 0$, $n_0 \equiv 1$ for $\lambda = 5/4$ and different values I are shown in Fig.1. For $\mu = \mu_* \equiv 5q'/(8A_0^2)$, corresponding to $\lambda = 1$, the FP coincide with initial soliton parameter $n_0 \equiv 1$, $y_0 = 0$. In this case soliton`s parameters are constant at time.

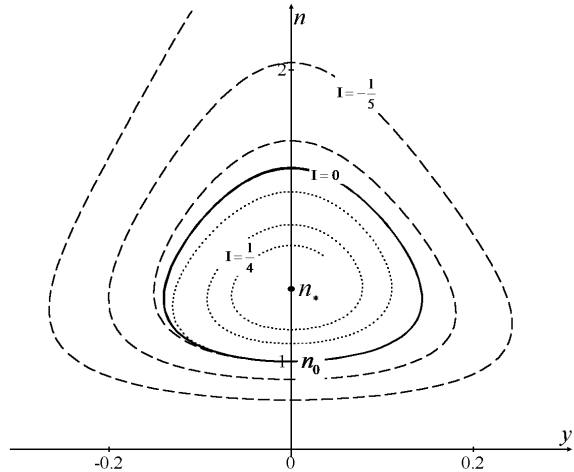

Fig. 1. The trajectories (8) on plane $(y, n)$ for $\lambda = 5/4$, initial conditions $y_0 = 0$, $n_0 \equiv 1$, and different values of I.

## 4. Numerical results

We now aim to solve the initial-value problem for the dynamics of the wave packet, $U(\xi, t=0) = \exp[i\phi(\xi)]\operatorname{sech}\xi$ with spatial phase distribution $\phi(\xi) = (3/2)\gamma\xi - (3/2)\gamma\tanh\xi + \gamma\ln(\cosh\xi)$ (corresponding to wavenumber $K(\xi) = d\phi/d\xi \equiv (3/2)\gamma\tanh^2\xi + \gamma\tanh\xi$), in the framework of Eq.(1), for $q(\xi) = 1 - \xi/10$ and different values of $\mu$ and $\gamma$ numerically. The analytically predicted equilibrium value of strength of the pseudo-SRS term from (8) for the initial pulse is $\mu_* = 1/16$. In direct simulations, the initial pulse for $\mu = 1/16$ and $I = 4\sqrt{3}\gamma \geq 0$ is transformed into a stationary localized distribution (Fig. 2.a). For $I < 0$ the initial pulse is unstable at time (Fig. 2.b).

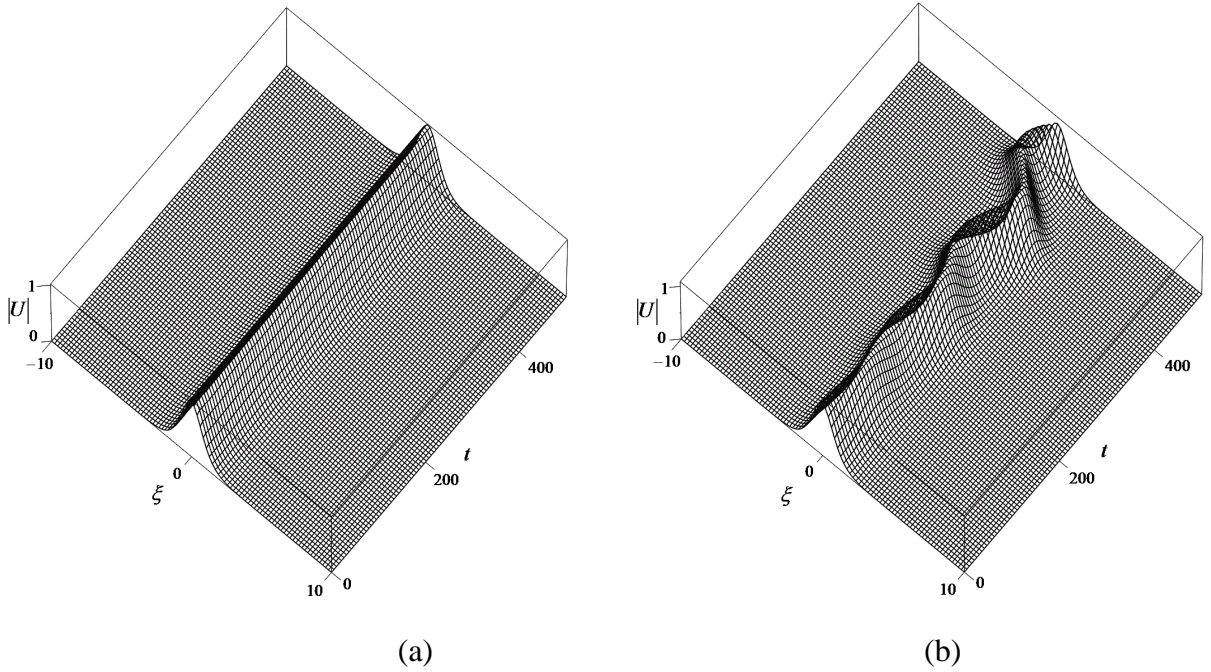

(a)          (b)

Fig.2. Numerical results for space-time distributions of $|U(\xi,t)|$ for $\mu = 1/16 \equiv \mu_*$ and different values of $I = 4\sqrt{3}\gamma$. [(a): $I \geq 0$, (b): $I = -1/4$].

Variation of the parameter $\mu$ leads to variation of the soliton's parameters (wavenumber and amplitude). Corresponding spatial distributions of $|U|$ at different moments of time for $\mu = 5/64 \equiv (5/4)\mu_*$ and different values of $I = 4\sqrt{3}\gamma$ are shown in Fig. 3.

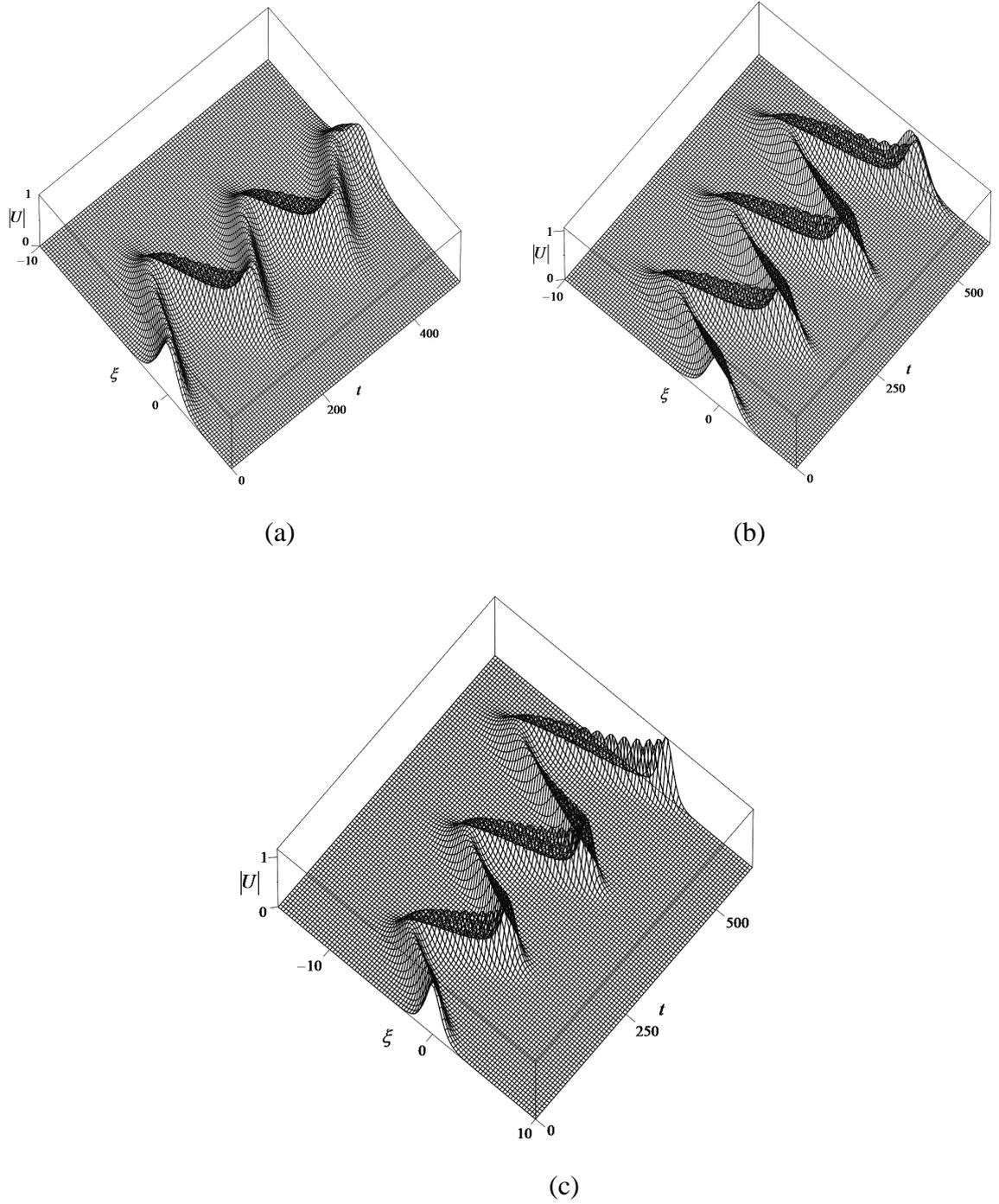

Fig.3. Numerical results for space-time distributions of $|U(\xi,t)|$ for $\mu = 5/64 \equiv (5/4)\mu_*$ and different values of $\mathrm{I} = 4\sqrt{3}\gamma$. [(a): $\mathrm{I}=1/4$, (b): $\mathrm{I}=0$, (c): $\mathrm{I}=-1/5$].

In figure 4, numerical results produced, as functions of time, by the simulations for the value of point coordinate of the maximum modulus of the wave-packet's shape $\xi_m$ ( $\max|U(\xi,t)|=|U(\xi_m,t)|$ ), are compared with the analytical counterparts of the mass-center wave-packet envelope $\bar{\xi} \equiv q_0(n-1)/q'$ obtained from Eqs. (6) for $\mu = (3/4)\mu_* \equiv 3/64$ and different values of $\mathrm{I}$.

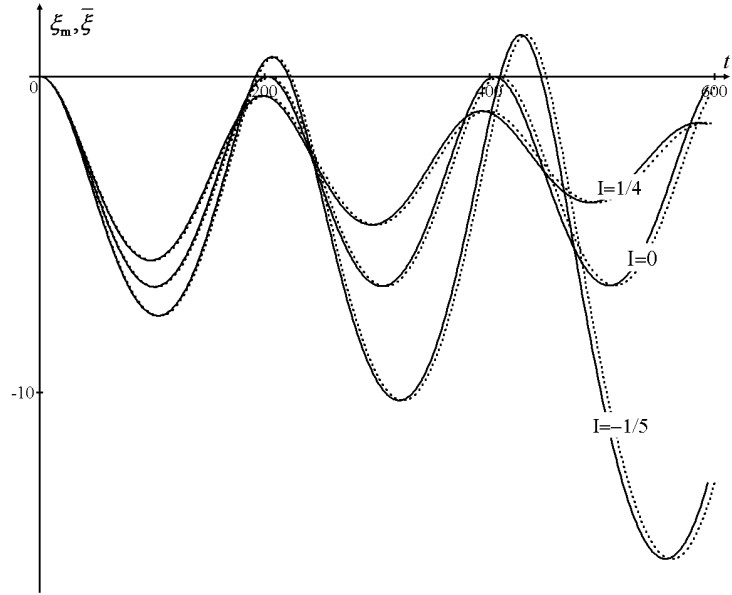

Fig. 4. Numerical results (solid curves) for the value of point coordinate of the maximum modulus of the wave-packet's shape $\xi_m$ and analytical results (dashed curves) for the mass-center wave-packet envelope $\bar{\xi}$ for $\mu = (3/4)\mu_* \equiv 3/64$, and different values of I.

## Conclusion

In this work, we studied the soliton dynamics in the framework of the extended inhomogeneous NLSE, includes the pseudo-SRS effect, the lineally decreasing SOD and TOD. The results were obtained by means of analytical method, based on evolution equations for the field moments, and verified by direct simulations. The stationary solitons exist due to the balance between the self-wavenumber downshift, caused by the pseudo-SRS, and the upshift induced by the decreasing SOD. The analytical solutions are close to their numerically found counterparts.

## Acknowledgements

This work was supported by the Russian Foundation for Basic Research projects No 15-02-01919 a.